\documentclass[11pt,a4paper]{article}

\usepackage{jheppub}
\usepackage{graphicx}
\usepackage{dcolumn}
\usepackage{bm}

\usepackage[utf8]{inputenc}
\usepackage[T1]{fontenc}
\usepackage{mathptmx} 

\usepackage{graphicx}
\usepackage{dcolumn}
\usepackage{bm}

\usepackage{booktabs} 
\usepackage{subcaption}
\usepackage{caption}
\begin{document}


\title{\boldmath Symmetric Dual Breather Modes as a Dynamic Origin of Tiny Particle Masses}

\author[a]{Yoritaka Iwata}
 \affiliation[a]{Osaka University of Economics and Law, \\
 Gakuonji 6-10, Osaka 581-0853, Japan}
\emailAdd{y-iwata@s.keiho-u.ac.jp}
\author[b]{Yasuhiro Takei}%
\affiliation[b]{ Mizuho Research \& Technologies, \\
Kanda Nishiki Cho 2-3, Tokyo 101-8443, Japan\\
}%



\date{\today}

\abstract{
Spontaneous symmetry breaking (SSB) is conventionally understood through static vacuum selection in low-energy approximations. However, the nonperturbative dynamical pathways connecting symmetric configurations to asymmetric states remain an open question. In this paper, we propose an operator-theoretic framework that characterizes $Z_{2}$ symmetry breaking in scalar field theory through spectral branching. Specifically, we identify an effective invariant manifold of the nonlinear Klein-Gordon equation termed the "symmetric dual breather mode," whose phase-space geometry is traced using a $t$-dependent finite-dimensional representation. Our analysis demonstrates that static SSB represents merely a small-amplitude limit ($A \ll \phi_{v1}$). As the amplitude increases into the heteroclinic regime ($A > \phi_{v1}$), the field dynamically overflows the central barrier, yielding a vanishing time-averaged order parameter ($\langle \phi \rangle = 0$) and driving a geometric curvature cancellation ($\langle V''(\phi) \rangle_{t} \to 0$). Governed by a universal scaling function $\mathfrak{F}(y)$, the logarithmic period divergence near the heteroclinic boundary naturally generates sub-eV effective mass scales without fine-tuning bare parameters. Furthermore, this dynamic barrier overflow naturally favors Majorana-like self-conjugate states over Dirac-type states. Additionally, nonlinear phase interference along the heteroclinic channel induces asymmetric phase-space contraction, decoupling the right-handed mode and dynamically giving rise to the left-handed Weyl equation. These results bridge nonperturbative wave dynamics, topological mass suppression, chiral selection, and the dynamic origin of active neutrino masses.
}

\maketitle


\section{\label{sec:level1} Introduction}

Spontaneous symmetry breaking (SSB) is a foundational concept in modern physics, describing how a system governed by symmetric laws can manifest in an asymmetric state~\cite{1961Goldstone,1964Higgs}. Conventionally visualized as a ``Mexican hat'' potential where the system spontaneously chooses a single degenerate vacuum~\cite{1996Weinberg}, this standard picture represents a static, low-energy approximation strictly corresponding to the small-amplitude limiting case. In the context of nonlinear wave equations, the full dynamical pathway from a symmetric to an asymmetric configuration requires a space-time dependent treatment beyond static energy minimization~\cite{Makhankov1978}.

In this paper, we characterize SSB in nonlinear wave systems as an operator-theoretic spectral branching~\cite{1966Kato}. Considering the nonlinear Klein-Gordon equation with a $Z_2$-symmetric double-well potential~\cite{1961Nambu}, we define symmetry breaking dynamically through the spectral evolution of the linearized operator. A singular symmetric state dynamically branches into two distinct, stable spectral sectors, connecting to bifurcation phenomena in nonlinear mathematics~\cite{1971Crandall,1983Guckenheimer,1981Henry, 1992Vanderbauwhede}. This ensures the $Z_2$ symmetry is realized as a dynamic, localized coexistence of phase-inverted field configurations, manifested in transient, spatially localized states termed ``symmetric dual breather modes.''

To capture this nonperturbative behavior, we employ a $t$-dependent finite-dimensional representation~\cite{AIP_TimeDependent_2026}, visualizing complex PDE dynamics as coupled rotational motions in a reduced phase space. As the field amplitude increases, the system enters a heteroclinic regime where the field dynamically overflows the central barrier, yielding a vanishing time-averaged order parameter ($\langle \phi \rangle = 0$).

This dynamical framework has profound implications for elementary particle physics.
While the ultra-light nature of neutrinos \cite{2008Gonzalez} is conventionally addressed through structural frameworks like the seesaw mechanism \cite{1977Minkowski,1979Yanagida}, our model offers a complementary, nonperturbative perspective.
By demonstrating that heteroclinic dynamics naturally favor Majorana-like states over Dirac-type states, we show that remarkably small mass scales and particle-antiparticle indistinguishability can emerge directly from the intrinsic geometric curvature cancellation of a single scalar field
without requiring explicit symmetry-breaking terms or heavy right-handed extensions.
This establishes a novel paradigm where the Majorana nature of active neutrinos is not an input parameter or a static field constraint, but rather a dynamic realization arising from topological limit cycles in phase space.

The remainder of this paper is organized as follows. Sec.~\ref{sec:theory} presents the theoretical formulation of $Z_2$ symmetry breaking and spectral branching. Sec.~\ref{sec:methods} specifies the numerical settings. Sec.~\ref{sec:results} establishes the dynamical regime classification, demonstrates dual breather emergence, and discusses neutrino physics implications. Conclusions are drawn in Sec.~\ref{sec:conclusion}.

\section{\label{sec:theory} Symmetry breaking in the scalar field} 
\subsection{Theoretical Model}
Spontaneous symmetry breaking (SSB) is a foundational mechanism for understanding nonperturbative field dynamics and particle mass generation. In this study, we consider the Lagrangian density of a real $\phi^4$-theory (for mathematical foundation, see \cite{12Jacobi}, and for physical foundation, see \cite{1965bjorken}):
\begin{equation} \label{lagrangian}
{\mathcal L} = \dfrac{1}{2} \partial_{\mu} \phi \partial^{\mu} \phi - V(\phi), 
\end{equation}
with the classical double-well potential
\begin{equation} \label{potential}
V(\phi) = \dfrac{1}{2} \mu^2 \phi^2 + \dfrac{1}{4} \lambda \phi^4, 
\end{equation}
where $\phi$ denotes a neutral scalar field, $\partial_{\mu} = (\partial_t, \nabla)$ is the four-gradient operator, $\mu^2 < 0$ is the mass parameter, and $\lambda > 0$ represents the self-interaction strength. The potential $V(\phi)$ belongs to the universality class of the Ising model~\cite{1963huang}.

The vacuum expectation values (VEVs) are the stationary points satisfying $\partial V / \partial \phi = 0$:
\begin{equation} \label{r-eq}
 \phi \left(\mu^2 + \lambda \phi^2 \right) = 0,
\end{equation}
yielding two symmetric broken vacua $\phi_{v1, v2} = \pm \sqrt{-\mu^2/\lambda}$ and an unstable central barrier peak $\phi_{v3} = 0$.
The spatio-temporal evolution is governed by the nonlinear Klein-Gordon equation:
\begin{equation} \label{eq-EL}
\partial_\mu \partial^\mu \phi + \mu^2 \phi + \lambda \phi^3 = 0,
\end{equation}
which intrinsically conserves total energy and momentum via Noether's theorem.

While standard quantum field formulations often employ complex fields in higher dimensions, restricting our analysis to a real-valued field in (1+1) dimensions isolates the purified dynamics of vacuum selection and barrier penetration.
This restriction provides a self-adjoint operator formalism ($\phi = \phi^\dagger$) within $L^2$ space~\cite{vonNeumann1932}. 
Unlike complex $U(1)$ fields where continuous phase rotations circumvent the central barrier, this self-adjoint constraint forces the real-valued wave dynamics to directly confront the tachyonic barrier $(\phi=0)$, laying the mathematical foundation for spectral branching and dynamic Majorana-like self-conjugacy.

\subsection{$Z_2$ Symmetry Breaking and Phase Coexistence}
Restricting $\phi$ to the real axis projects the continuous $U(1)$ symmetry down to a discrete $Z_2$ parity symmetry ($\phi \to -\phi$). In this setting, opposite-phase states correspond to sign-inverted configurations $\phi(x,t)$ and $-\phi(x,t)$. 

Under static SSB, the system selects either $\phi_{v1}$ or $\phi_{v2}$ as a uniform ground state, separated by an energetic potential barrier (Fig.~\ref{fig:schematic_regimes}). While static topological kinks interpolate between $\phi_{v1}$ and $\phi_{v2}$ across space, non-topological configurations allow opposite-phase sectors to dynamically interact and oscillate within a shared spatial domain. This raises a fundamental question: \textit{Under what mathematical conditions can the localized coexistence of opposite-phase states be dynamically sustained as stable, non-radiating oscillations beyond simple static kinks?}

\subsection{Physical Implications for Parity Coexistence and CP Violation}
The projection of complex phase dynamics into a binary $Z_2$ vacuum choice provides a geometric lens for discrete symmetries. While the dual breather mode (formulated below) preserves overall spatial parity, its underlying operator-theoretic spectral branching enables localized phase coexistence. 

This mechanism suggests that phenomenological Charge-Parity (CP) violation and emergent Majorana properties in neutrino physics can be reinterpreted as dynamical consequences of confining complex phase degrees of freedom into real-valued, branched spectral sectors. Parity symmetry is maintained at the structural operator level while transferring phase information into physical orbit dynamics, providing a nonperturbative foundation for neutrino mass generation without introducing static axial backgrounds.

\subsection{Breather Mechanism and Spectral Branching}
The classical single breather solution of Eq.~\eqref{eq-EL} represents a localized, time-periodic bound state oscillating around a single vacuum $\phi_{v i} \in \{\phi_{v1}, \phi_{v2}\}$~\cite{1982Rajaraman, 1983Campbell}:
\begin{equation} \label{breather-sol}
\phi_{B i}(x,t) = \phi_{v i} \frac{\sqrt{1-\omega^2} \cos(\omega t)}{\cosh(x \sqrt{1-\omega^2}) + \omega \cos(\omega t)},
\end{equation}
where $0 < \omega < 1$ is the internal frequency. 
Although $\phi^4$ theory lacks infinitely persistent non-radiating single breathers~\cite{1987Segur, 1991Boyd}, higher-order dynamical configurations can achieve structural stabilization.

We propose the \textit{symmetric dual breather mode}, defined as a parity-symmetric superposition:
\begin{equation}
\phi_{\text{dual}}(x,t) = \phi_{B 1}(x,t) + \phi_{B 2}(-x,t).
\end{equation}
This state describes simultaneous, phase-inverted oscillations around both vacuum sectors. Rather than undergoing global uniform symmetry breaking, the field exhibits localized spectral branching where energy concentrates into specific parity-symmetric sectors. 

In this framework, the term ``breather mechanism'' broadly encompasses the nonlinear dynamics wherein pulsating dual breather configurations mediate transitions between static symmetry breaking, dynamic curvature cancellation, and the heteroclinic emergence of sub-eV mass scales.

\subsection{Symmetric Dual Breather Mode as an Effective Invariant Manifold}
The central localized excitation arising from this operator-theoretic framework is the symmetric dual breather mode. Rather than a transient numerical artifact, this state represents a robust, parameter-dependent effective \textit{invariant manifold} in phase space.

Structurally, the dual breather balances out-of-phase configurations ($\phi(x,t) \approx -\phi(-x,t)$) across the central barrier. Along this invariant manifold, the field undergoes persistent heteroclinic motion bridging the two symmetric $Z_2$ sectors without dissipative decay. By organizing global phase space into distinct flow regime, this mode forms the nonperturbative backbone for dynamic barrier overflow and the resulting time-averaged mass cancellation $m_{\text{eff}}^2 \propto \langle V''(\phi) \rangle_t \to 0$ at the leading order.

\begin{figure}[tb]
\centering
\includegraphics[width=0.6\linewidth]{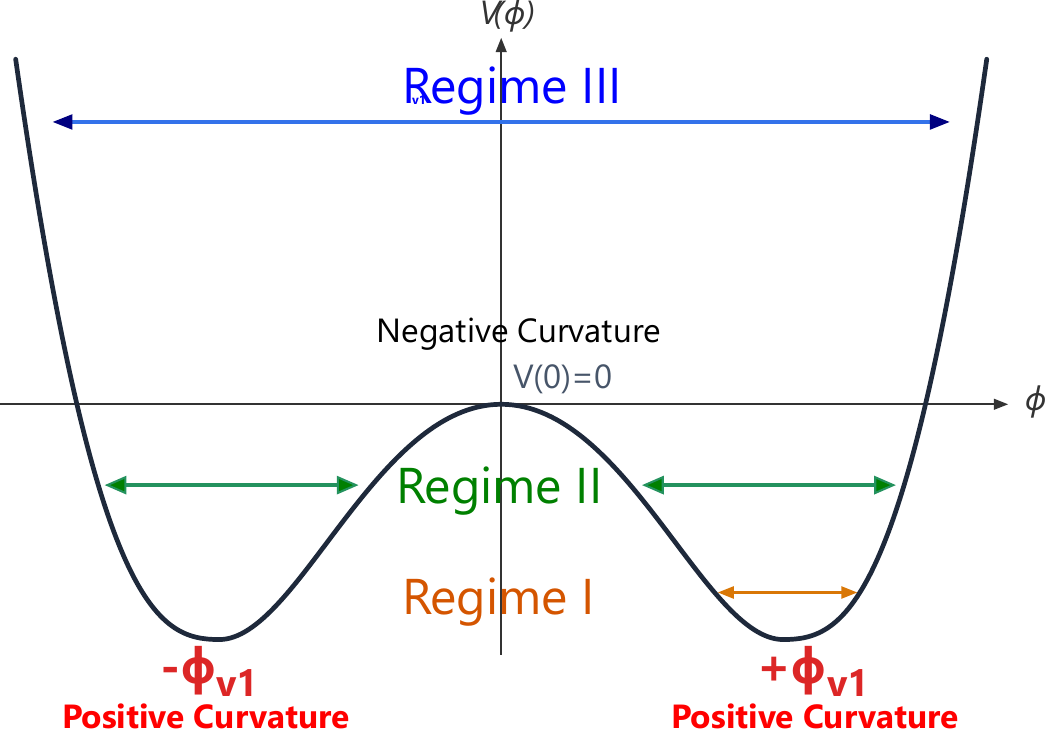}
\caption{\textbf{Spatio-temporal field dynamics across distinct energy regimes.}
(a--e) Regime I ($A = 0.2 \ll \phi_{v1}$): Microscopic harmonic oscillations localized near a single vacuum ($\phi \approx \phi_{v1}$). 
(f--j) Regime II ($A = 1.0 \approx \phi_{v1}$): Persistent spatio-temporal wave packet of the symmetric dual breather mode, exhibiting symmetric oscillations around the positive and negative vacua (note that the breather amplitude remains very small in this parameter setting).
(k--o) Regime III ($A = 1.8 > \phi_{v1}$): Large-amplitude heteroclinic barrier-overflow dynamics continuously traversing between $\pm\phi_{v1}$.
In each case, light blue curves depict recent past profiles for visual clarity.
The horizontal axis denotes the spatial coordinate $x \in [0, 8]$ and the vertical axis represents the field amplitude $\phi \in [-2, 2]$ for (a-j), and $x \in [0, 8]$ and $\phi \in [-4, 4]$ for (k-o). }
\label{fig:schematic_regimes}
\end{figure}

\begin{table*}[tb]
\caption{\textbf{Sign sectors and physical characterization of the potential curvature $V''(\phi)$.} The domain is divided at the inflection points $|\phi| = \phi_{v1}/\sqrt{3}$.}
\label{tab:curvature_sectors}
\centering
\begin{tabular}{lll}
\hline\hline
Sector & Field Range $|\phi|$ & Curvature $V''(\phi)$ and Physical Behavior\\
\hline
Positive Curvature & $|\phi| > \phi_{v1}/\sqrt{3}$ & $V''(\pm\phi_{v1}) = -2\mu^2 > 0$: \\ & & 
Conventional restoring force near vacua $\pm\phi_{v1}$ \\
Inflection Point   & $|\phi| = \phi_{v1}/\sqrt{3}$ & $V''(\phi) = 0$: \\ & & 
Boundary marking the onset of instability \\
Negative Curvature & $|\phi| < \phi_{v1}/\sqrt{3}$ & $V''(0) = \mu^2 < 0$: \\ & & 
Local instability near central barrier ($\phi = 0$) \\
\hline\hline
\end{tabular}
\end{table*}

\section{\label{sec:methods} Methods}
\subsection{Precise Computing Algorithm}
To investigate the long-term stability and fine structure of the symmetric dual breather mode, we employ a high-precision numerical scheme that rigorously conserves the Hamiltonian and $Z_2$ symmetry. Time integration is performed using the Gauss-Legendre implicit Runge-Kutta method~\cite{1964Butcher,87Butcher, 20Rana, 07Mardal, 06Staff,1993Hairer}, celebrated for its high-order accuracy and symplectic properties~\cite{2006Hairer} (for application to partial differential equations, see \cite{Isabel}). Spatial discretization utilizes the Fourier spectral method~\cite{86Canuto, 06canuto, 97Gottlieb,01Boyd}, exploiting exponential convergence for periodic domains.

This combined algorithm maintains relative energy errors within $\mathcal{O}(10^{-14})$ throughout long-term simulations~\cite{20takeiiwata, 21takeiiwata-2, 20iwata-2, 21takeiiwata-3}. 
Such fidelity is essential for cleanly separating transient quasi-breather radiation from the structural stabilization of the invariant manifold.

\subsection{Numerical Settings and Regime Initialization}
Simulations are conducted in a 1D spatial domain $x \in [0, L]$ ($L = 8$) with periodic boundary conditions, eliminating artificial boundary reflections. We adopt natural units ($c = \hbar = 1$) with canonical potential parameters:
\begin{equation}
\mu^2 = -1.0, \quad \lambda = 0.5,
\end{equation}
which set the VEVs and potential inflection boundaries ($V''(\phi_c) = 0$) to:
\begin{equation}
\phi_{v1,v2} = \pm \sqrt{\frac{-\mu^2}{\lambda}} = \pm \sqrt{2} \approx \pm 1.414, \quad \phi_c = \frac{\phi_{v1}}{\sqrt{3}} \approx 0.816.
\end{equation}
In the small-amplitude limit around a single vacuum, this yields the static mass $m_{\text{static}}^2 = V''(\phi_{v1}) = 2.0$.

The initial profile is prescribed as a fundamental Fourier spatial mode:
\begin{equation} \label{init-cond}
\phi(x, 0) = A \sin\left(\frac{\pi x}{4}\right) + \phi_0,
\end{equation}
with zero initial velocity, $\partial_t \phi(x,0) = 0$, where $\phi_0$ specifies the background VEV shift and initial amplitude $A$ serves as the control parameter governing injected kinetic energy. Systematically tuning $A$ accesses three fundamental dynamical regimes (Fig.~\ref{fig:schematic_regimes}):

\subsubsection{Regime I ($A = 0.2 \ll \phi_c$, $\phi_0 = \phi_{v1}$)}
Microscopic oscillations trapped within a single potential well ($\phi \approx \phi_{v1}$), breaking spatial reflection symmetry to recover standard static SSB with $m_{\text{eff}}^2 \approx m_{\text{static}}^2 = 2.0$.

\subsubsection{Regime II ($A = 1.0 \approx \phi_c$, $\phi_0 = 0$)}
Intermediate-amplitude excitations initialized around the symmetric background ($\phi_0 = 0$), reaching the inflection threshold $\phi_c$ to drive nonlinear harmonic coupling that forms stable, spatial-parity-symmetric dual breather modes with suppressed effective mass ($0 < m_{\text{eff}}^2 \ll 2.0$).

\subsubsection{Regime III ($A = 1.8 > \phi_{v1}$, $\phi_0 = \phi_{v1}$)}
Large-amplitude heteroclinic overflow initialized with a background shift, continuously traversing across the central barrier ($\phi = 0$) between both vacuum sectors, realizing complete dynamic curvature cancellation $m_{\text{eff}}^2 \propto \langle V''(\phi) \rangle_t \to 0$ and emergent Majorana properties.

\subsection{Time-dependent Finite-dimensional Representation}
Following Ref.~\cite{AIP_TimeDependent_2026}, the infinite-dimensional PDE dynamics is projected onto a sequence of $t$-dependent, finite-dimensional phase-space snapshots. Evaluating ordinary differential equation (ODE) trajectories at each instantaneous snapshot $t$ reconstructs field dynamics as traceable geometric orbits in reduced phase space~\cite{74hirsch, 97temam}.

While single breathers map to localized orbits around a single vacuum, the symmetric dual breather mode manifests as counter-rotating, nonlinearly coupled orbits encircling both VEVs ($\phi_{v1}$ and $\phi_{v2}$). This finite-dimensional mapping provides an intuitive visual representation of phase-space bifurcations, limit cycles, and dynamic curvature cancellation.

\begin{figure*}[htbp]
\bf{Spatio-temporal field profile in Regime I.} \\
  \centering
  \begin{subfigure}[b]{0.19\textwidth}
    \centering
    \includegraphics[width=\textwidth]{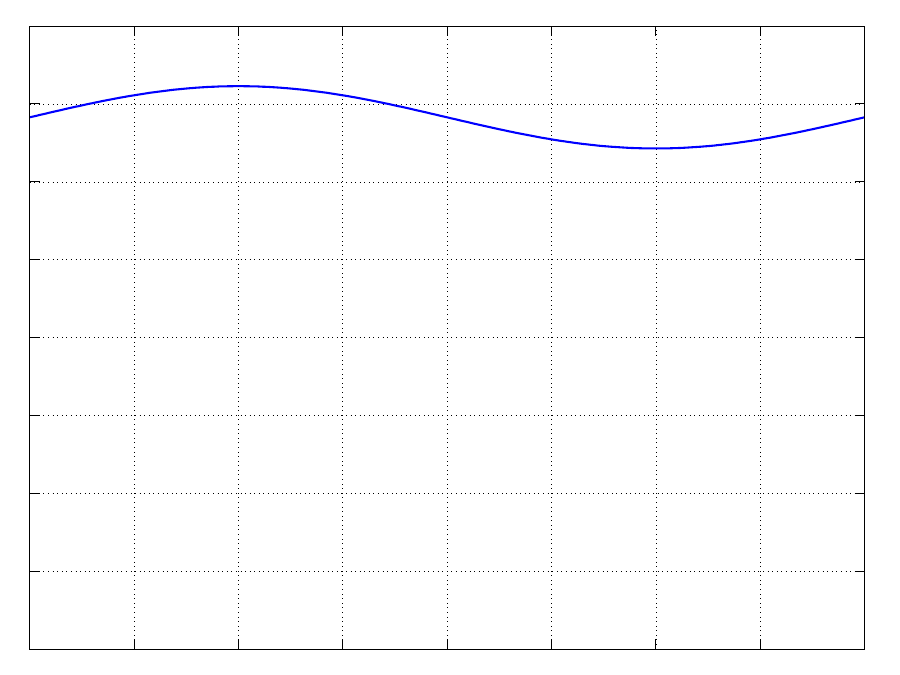}
    \caption{$t = t_0$}
  \end{subfigure}
  \hfill
  \begin{subfigure}[b]{0.19\textwidth}
    \centering
    \includegraphics[width=\textwidth]{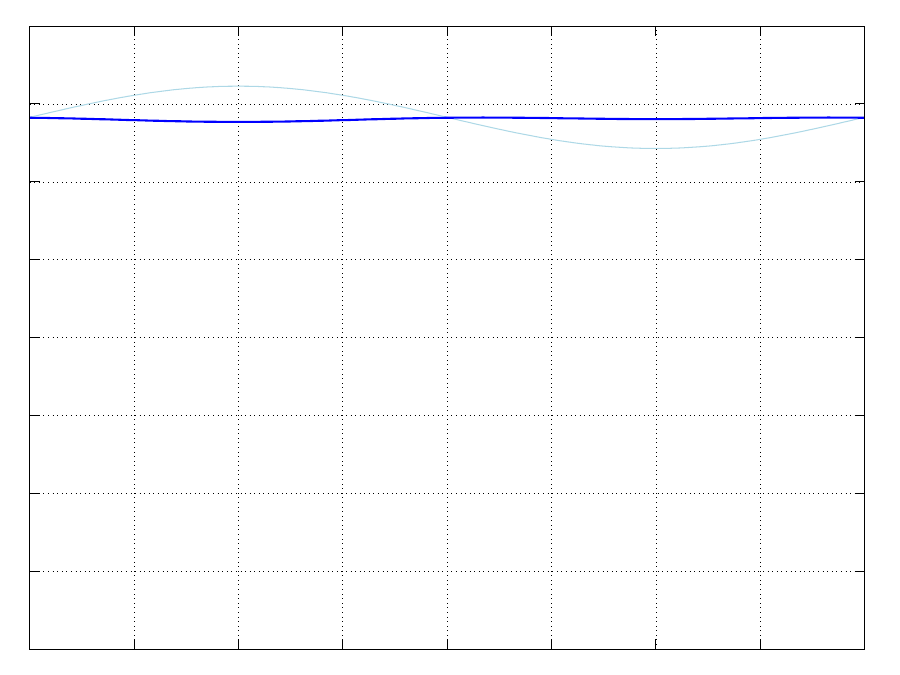}
    \caption{$t = t_1$}
  \end{subfigure}
  \hfill
  \begin{subfigure}[b]{0.19\textwidth}
    \centering
    \includegraphics[width=\textwidth]{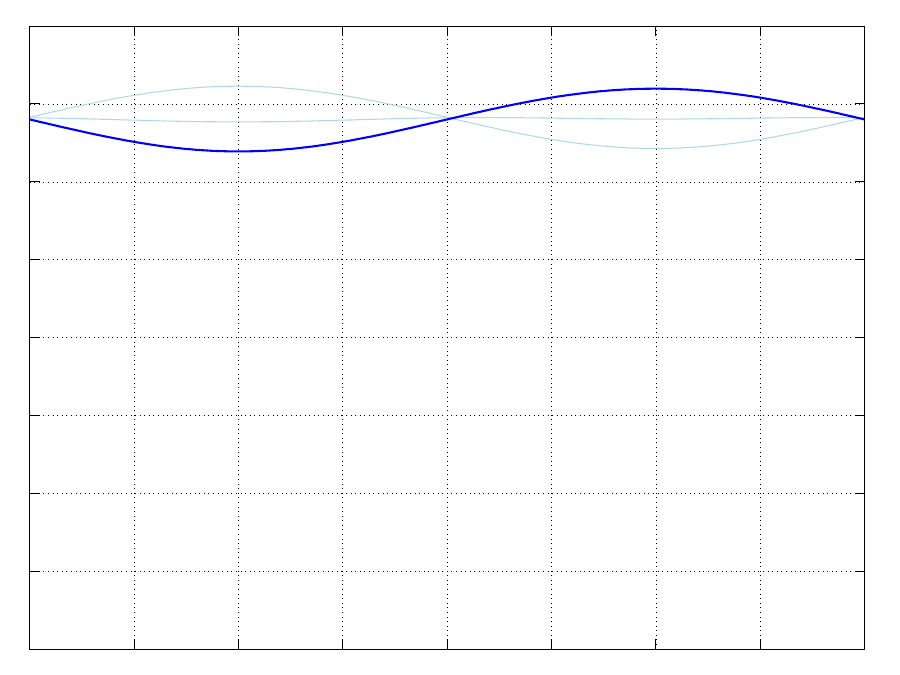}
    \caption{$t = t_2$}
  \end{subfigure}
  \hfill
  \begin{subfigure}[b]{0.19\textwidth}
    \centering
    \includegraphics[width=\textwidth]{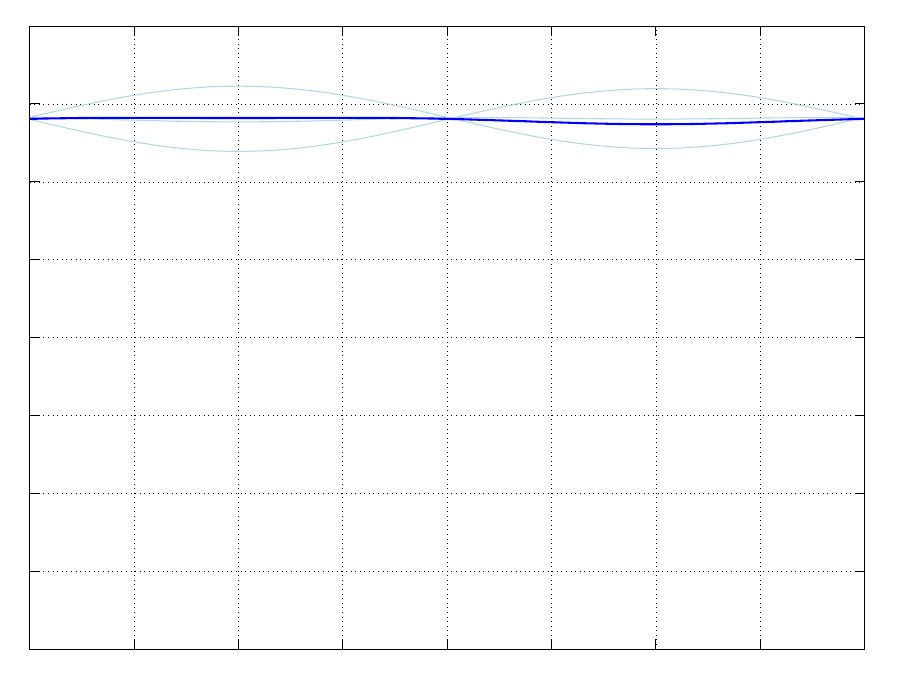}
    \caption{$t = t_3$}
  \end{subfigure}
  \hfill
  \begin{subfigure}[b]{0.19\textwidth}
    \centering
    \includegraphics[width=\textwidth]{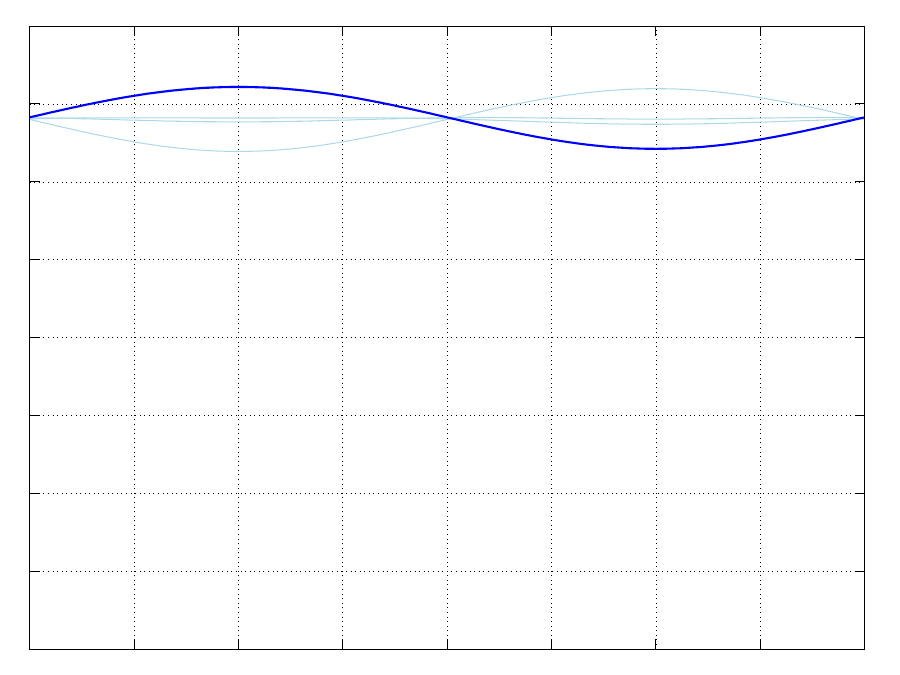}
    \caption{$t = t_4$}
  \end{subfigure}

  \bf{  \vspace{3mm} 
  Spatio-temporal field profile in Regime II.} \\
  \centering
  \begin{subfigure}[b]{0.19\textwidth}
    \centering
    \includegraphics[width=\textwidth]{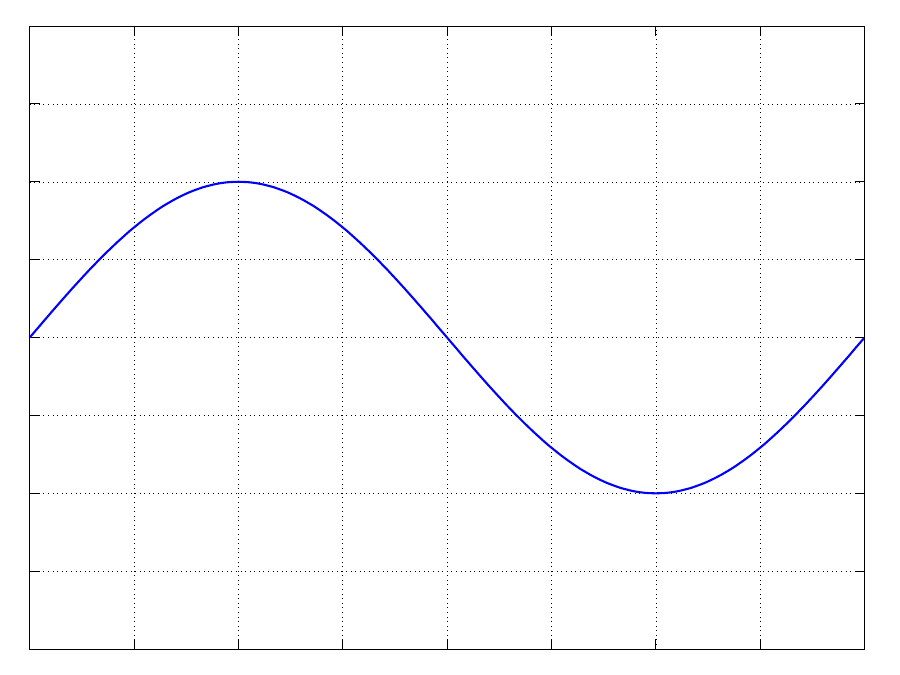}
    \caption{$t = t_0$}
  \end{subfigure}
  \hfill
  \begin{subfigure}[b]{0.19\textwidth}
    \centering
    \includegraphics[width=\textwidth]{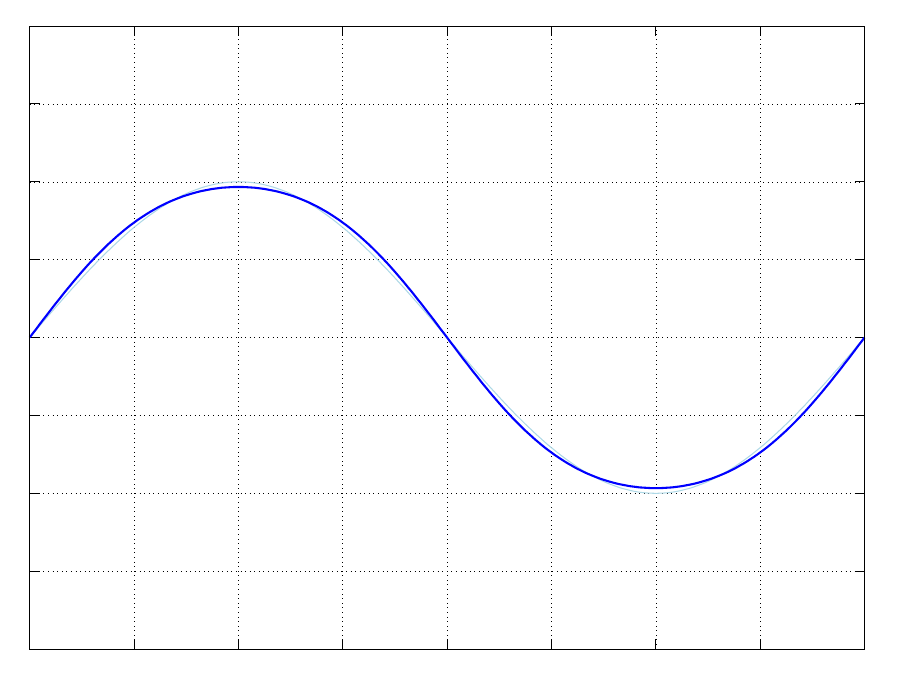}
    \caption{$t = t_1$}
  \end{subfigure}
  \hfill
  \begin{subfigure}[b]{0.19\textwidth}
    \centering
    \includegraphics[width=\textwidth]{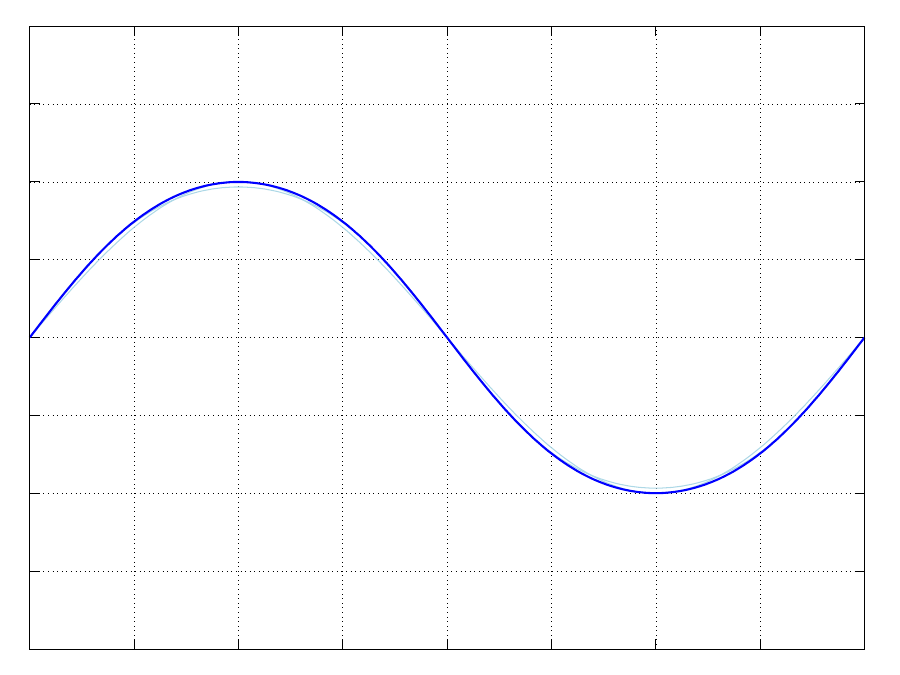}
    \caption{$t = t_2$}
  \end{subfigure}
  \hfill
  \begin{subfigure}[b]{0.19\textwidth}
    \centering
    \includegraphics[width=\textwidth]{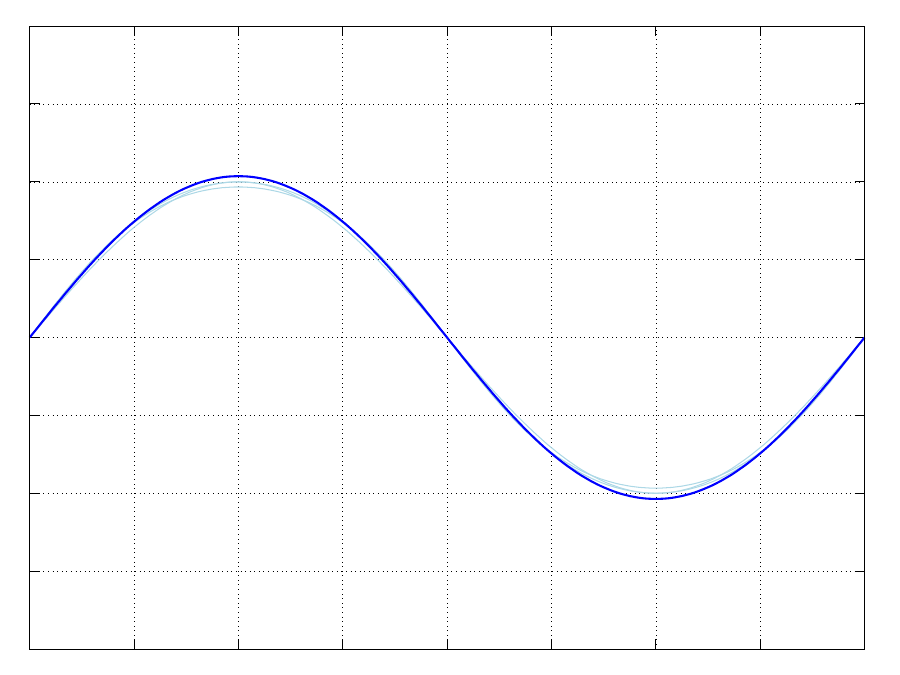}
    \caption{$t = t_3$}
  \end{subfigure}
  \hfill
  \begin{subfigure}[b]{0.19\textwidth}
    \centering
    \includegraphics[width=\textwidth]{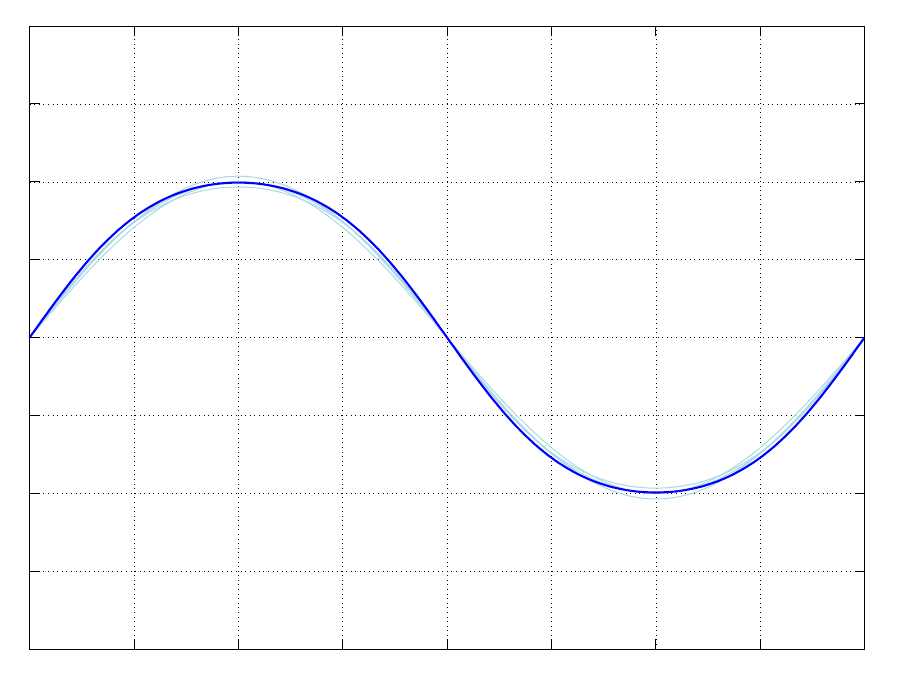}
    \caption{$t = t_4$}
  \end{subfigure}

\bf{ \vspace{3mm} 
Spatio-temporal field profile in Regime III.} \\
  \centering
  \begin{subfigure}[b]{0.19\textwidth}
    \centering
    \includegraphics[width=\textwidth]{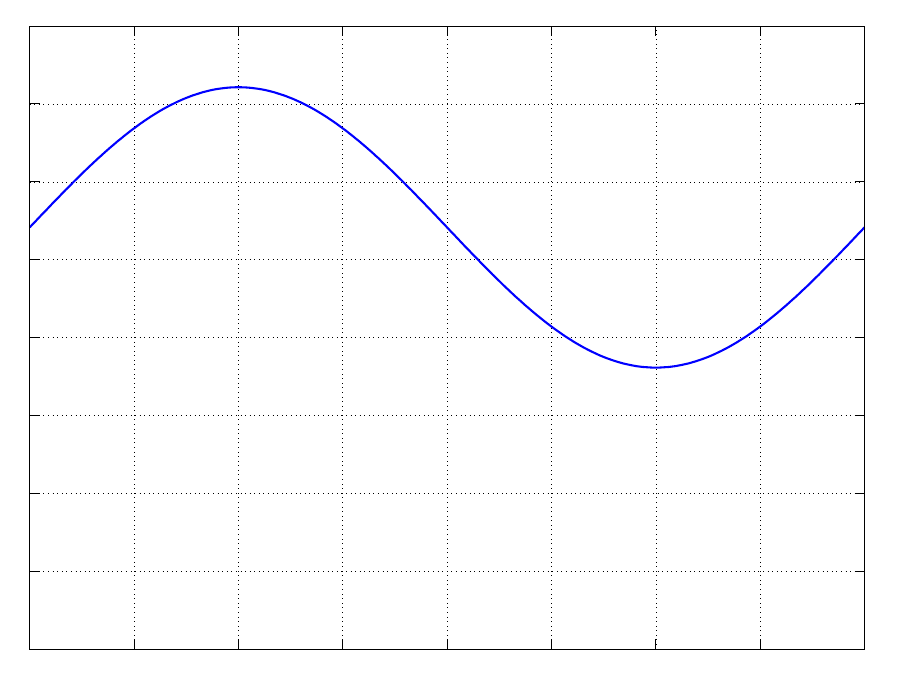}
    \caption{$t = t_0$}
  \end{subfigure}
  \hfill
  \begin{subfigure}[b]{0.19\textwidth}
    \centering
    \includegraphics[width=\textwidth]{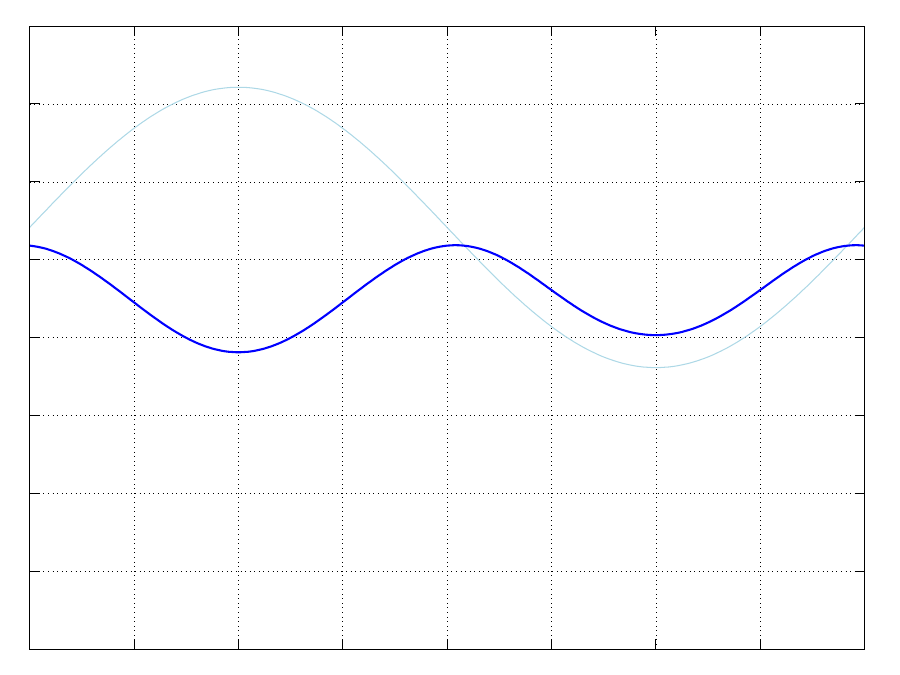}
    \caption{$t = t_1$}
  \end{subfigure}
  \hfill
  \begin{subfigure}[b]{0.19\textwidth}
    \centering
    \includegraphics[width=\textwidth]{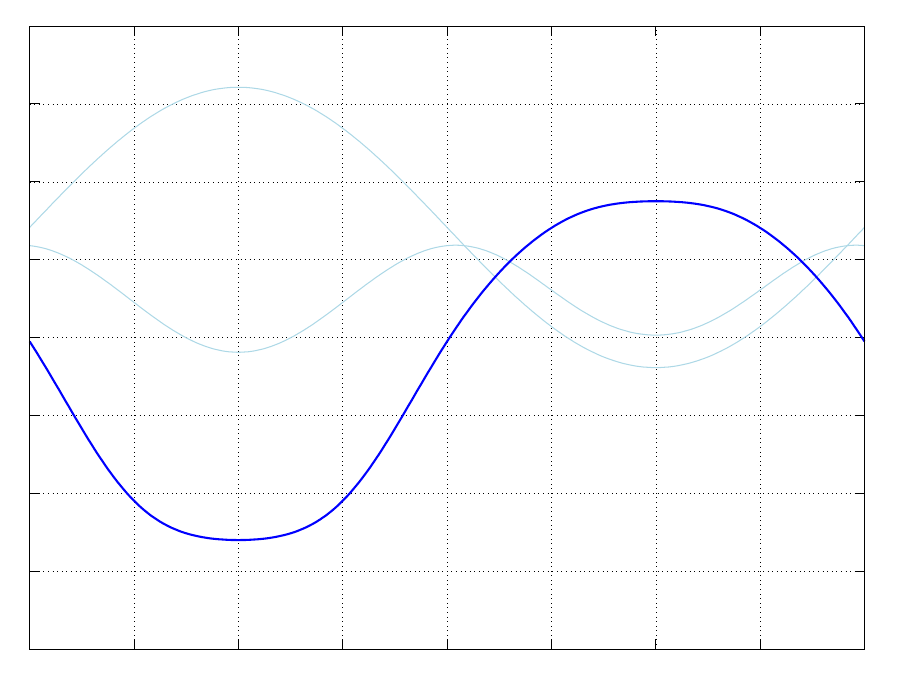}
    \caption{$t = t_2$}
  \end{subfigure}
  \hfill
  \begin{subfigure}[b]{0.19\textwidth}
    \centering
    \includegraphics[width=\textwidth]{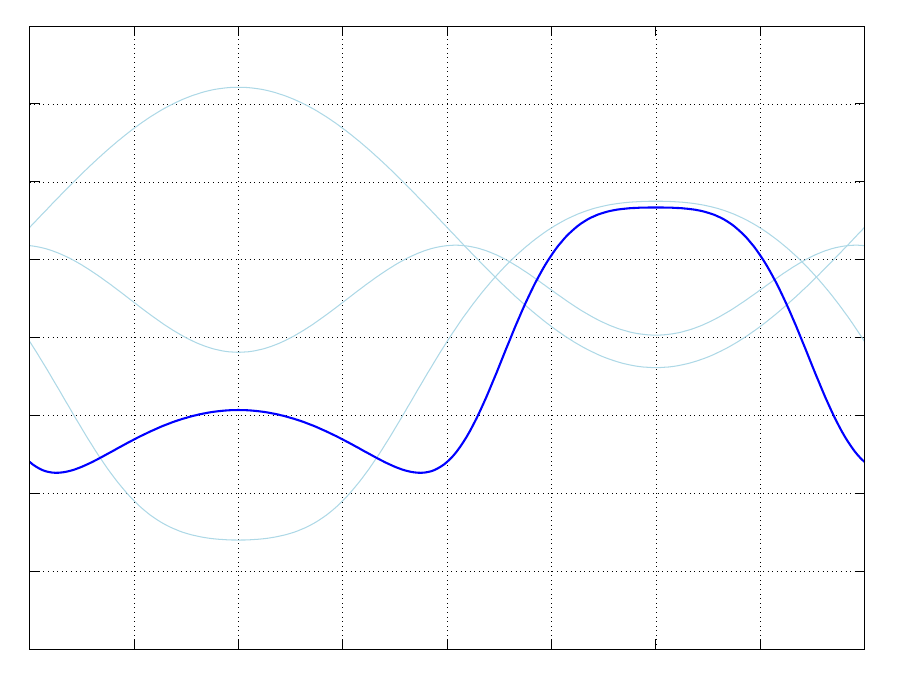}
    \caption{$t = t_3$}
  \end{subfigure}
  \hfill
  \begin{subfigure}[b]{0.19\textwidth}
    \centering
    \includegraphics[width=\textwidth]{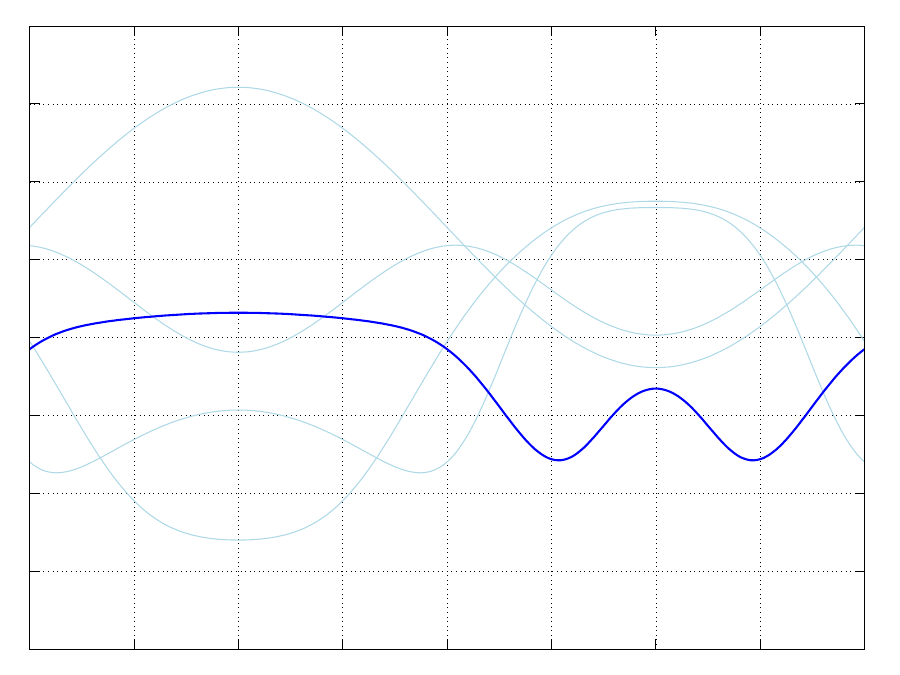}
    \caption{$t = t_4$}
  \end{subfigure}

 \caption{\textbf{Spatio-temporal field dynamics across distinct energy regimes.}
(a–e) Regime I ($A = 0.2 \ll \phi_{v1}$): Microscopic harmonic oscillations localized near a single vacuum ($\phi \approx \phi_{v1}$). 
(f–j) Regime II ($A = 1.0 \approx \phi_{v1}$): Persistent spatio-temporal wave packet of the symmetric dual breather mode, exhibiting symmetric oscillations around the positive and negative vacua (note that the breather amplitude remains very small in this parameter setting).
(k–o) Regime III ($A = 1.8 > \phi_{v1}$): Large-amplitude heteroclinic barrier-overflow dynamics continuously traversing between $\pm\phi_{v1}$.
In each case, light blue curves depict recent past profiles for visual clarity.
The horizontal axis denotes the spatial coordinate $x \in [0, 8]$ and the vertical axis represents the field amplitude $\phi \in [-2, 2]$ for (a-j), and $x \in [0, 8]$ and $\phi \in [-4, 4]$ for (k-o). }
\label{fig:field_profiles}
  \label{fig:regime_I_profile}
\end{figure*}

\begin{figure*}[t]
\captionsetup[sub]{justification=centering}

\centerline{\textbf{Finite-dimensional dynamical system representation in Regime I.}} 
\vspace{1.5mm}
\centering
\begin{subfigure}[b]{0.19\textwidth}
  \centering
  \includegraphics[width=\textwidth]{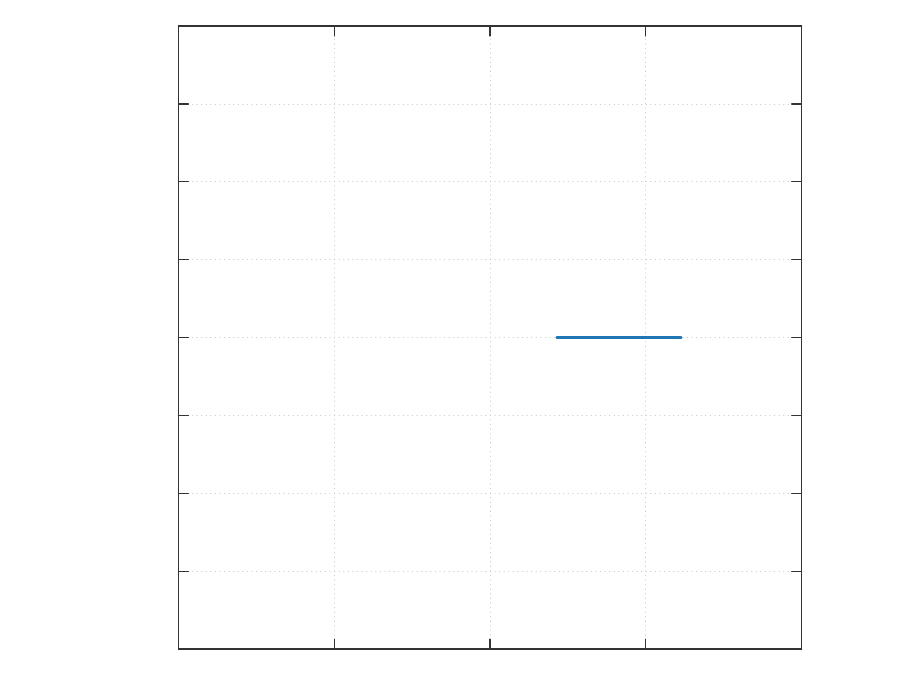}
 \caption{$t = t_0$}
\end{subfigure}
\hfill
\begin{subfigure}[b]{0.19\textwidth}
  \centering
  \includegraphics[width=\textwidth]{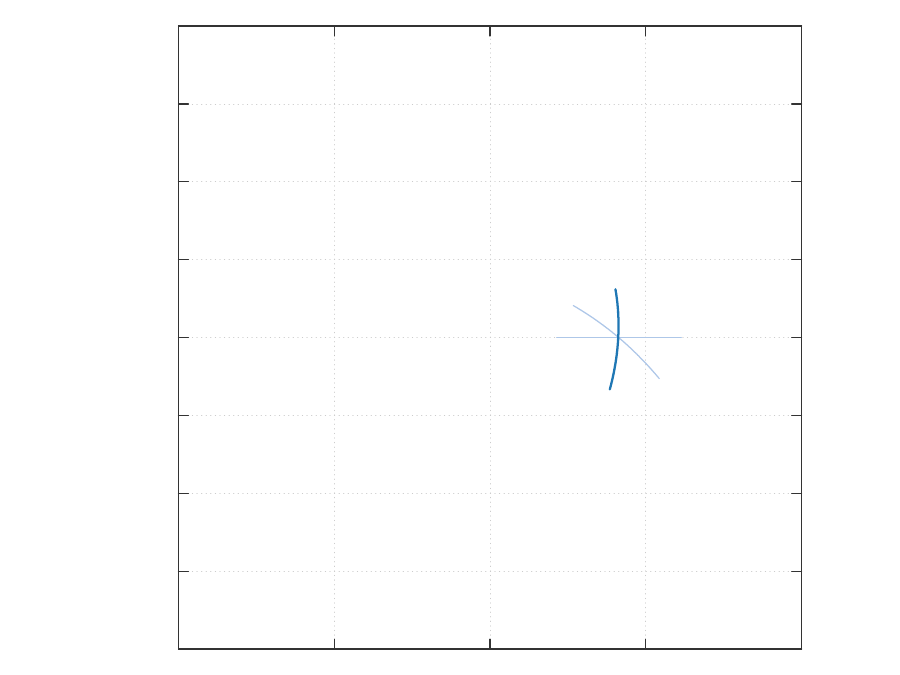}
  \caption{$t = t_1$}
\end{subfigure}
\hfill
\begin{subfigure}[b]{0.19\textwidth}
  \centering
  \includegraphics[width=\textwidth]{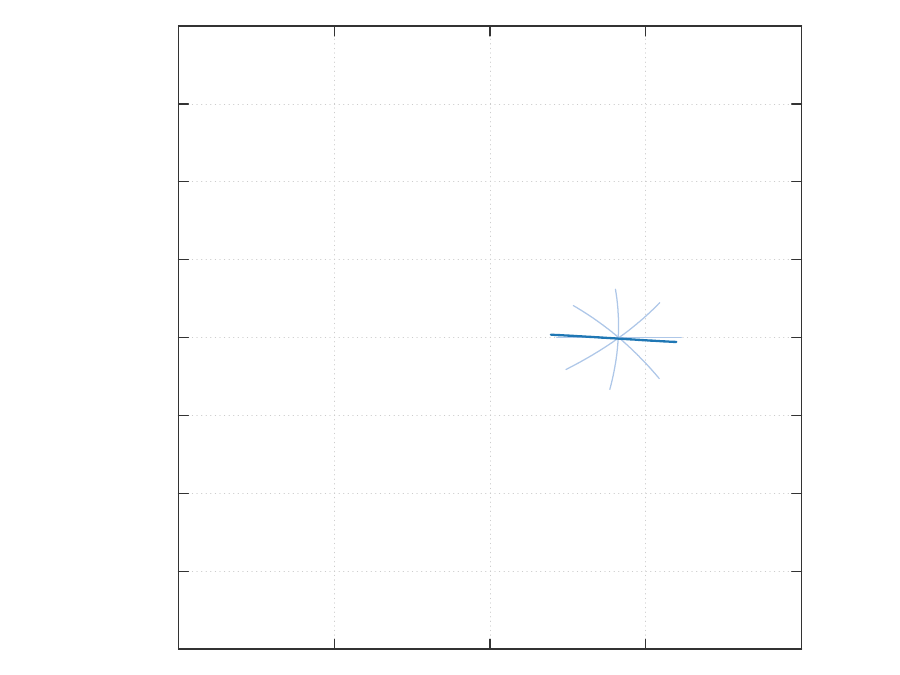}
  \caption{$t = t_2$}
\end{subfigure}
\hfill
\begin{subfigure}[b]{0.19\textwidth}
  \centering
  \includegraphics[width=\textwidth]{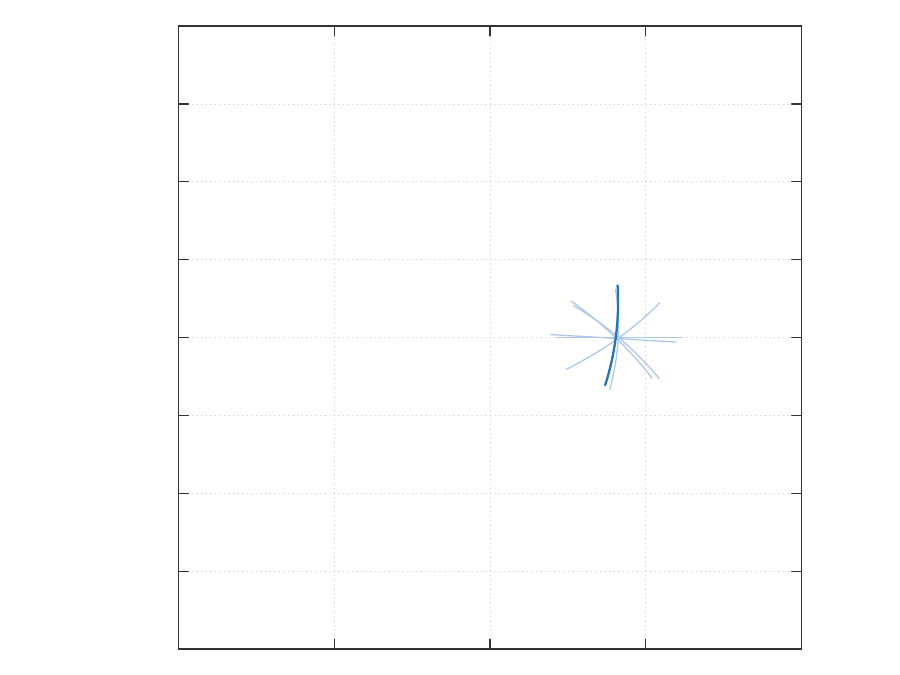}
  \caption{$t = t_3$}
\end{subfigure}
\hfill
\begin{subfigure}[b]{0.19\textwidth}
  \centering
  \includegraphics[width=\textwidth]{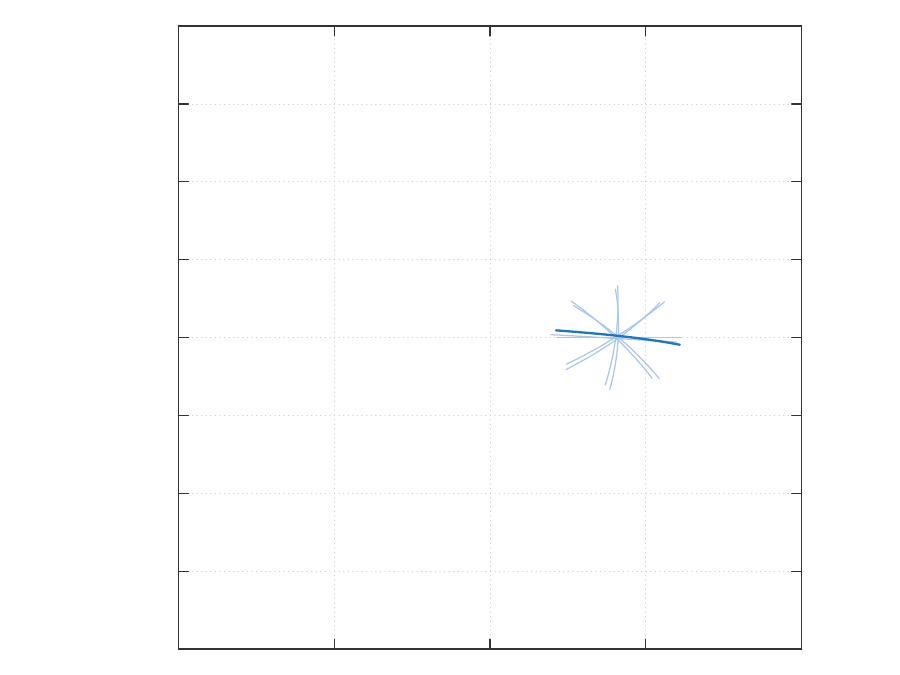}
  \caption{$t = t_4$}
\end{subfigure}

\vspace{3mm}
\centerline{\textbf{Finite-dimensional dynamical system representation in Regime II.}}
\vspace{1.5mm}
\centering
\begin{subfigure}[b]{0.19\textwidth}
  \centering
  \includegraphics[width=\textwidth]{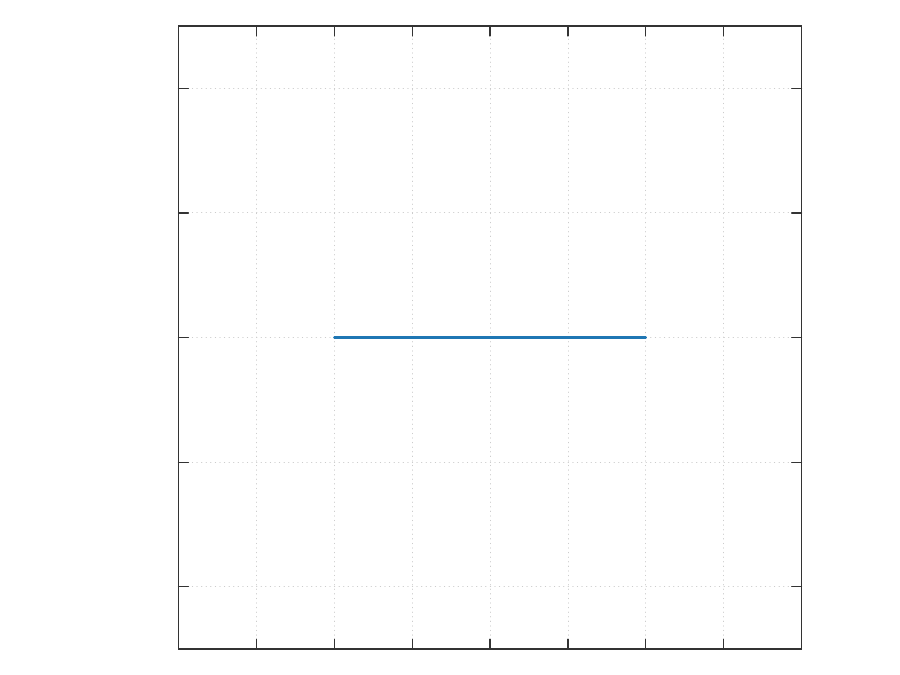}
  \caption{$t = t_0$}
\end{subfigure}
\hfill
\begin{subfigure}[b]{0.19\textwidth}
  \centering
  \includegraphics[width=\textwidth]{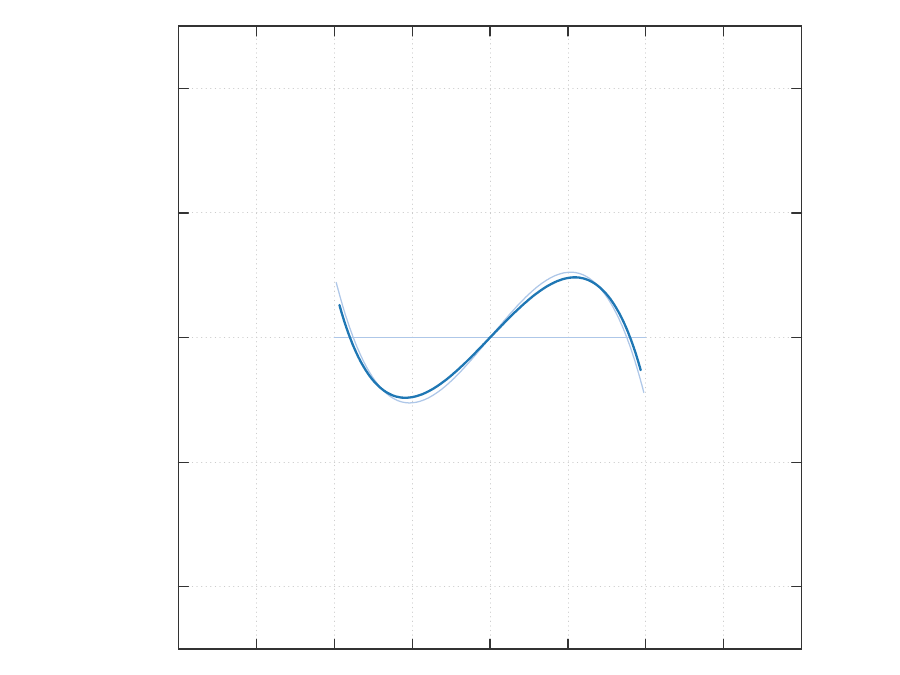}
  \caption{$t = t_1$}
\end{subfigure}
\hfill
\begin{subfigure}[b]{0.19\textwidth}
  \centering
  \includegraphics[width=\textwidth]{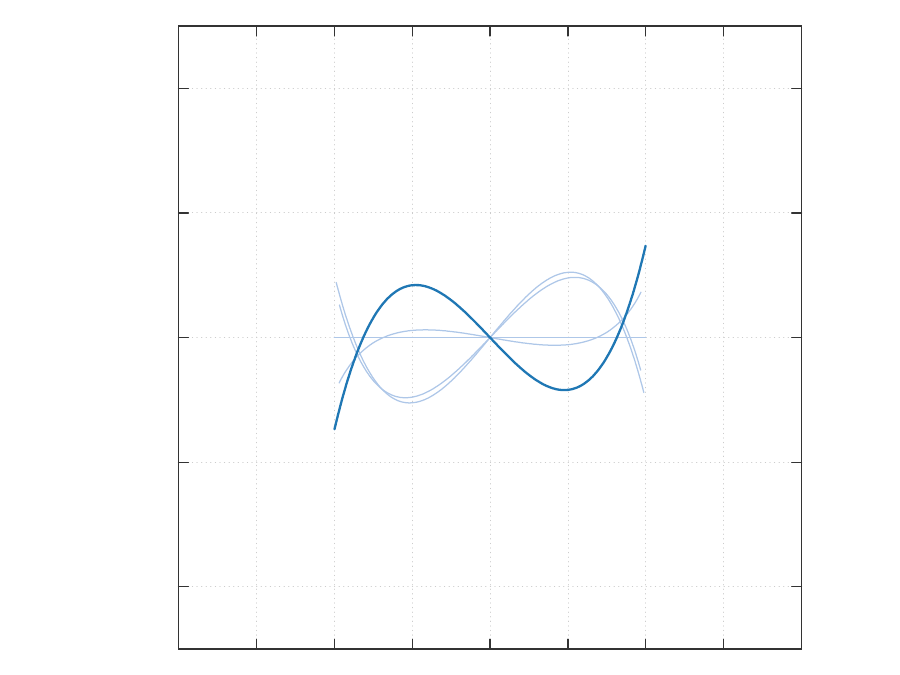}
  \caption{$t = t_2$}
\end{subfigure}
\hfill
\begin{subfigure}[b]{0.19\textwidth}
  \centering
  \includegraphics[width=\textwidth]{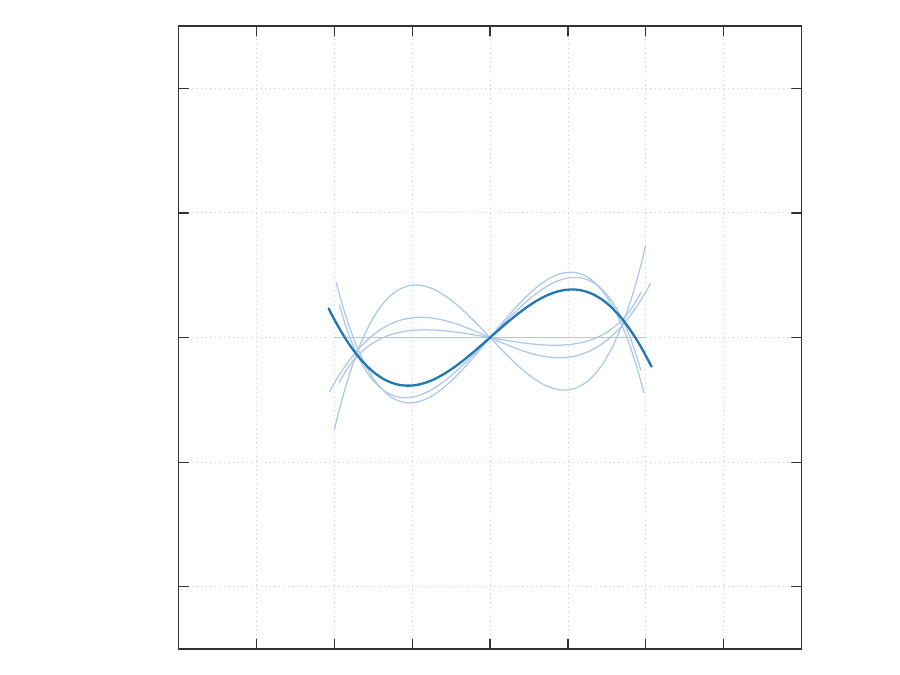}
  \caption{$t = t_3$}
\end{subfigure}
\hfill
\begin{subfigure}[b]{0.19\textwidth}
  \centering
  \includegraphics[width=\textwidth]{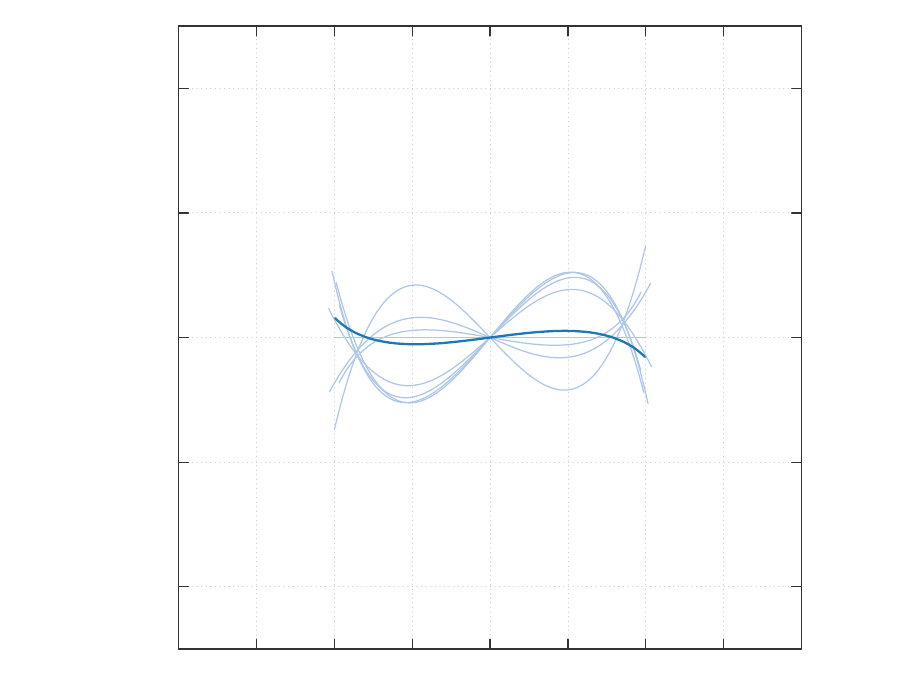}
  \caption{$t = t_4$}
\end{subfigure}

\vspace{3mm}
\centerline{\textbf{Finite-dimensional dynamical system representation in Regime III.}}
\vspace{1.5mm}
\centering
\begin{subfigure}[b]{0.19\textwidth}
  \centering
  \includegraphics[width=\textwidth]{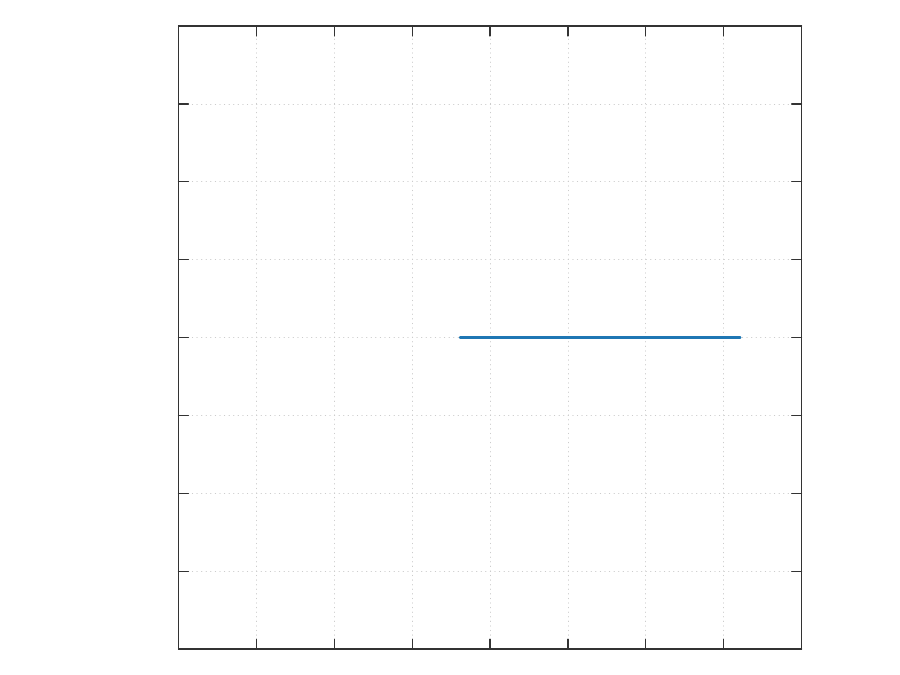}
  \caption{$t = t_0$}
\end{subfigure}
\hfill
\begin{subfigure}[b]{0.19\textwidth}
  \centering
  \includegraphics[width=\textwidth]{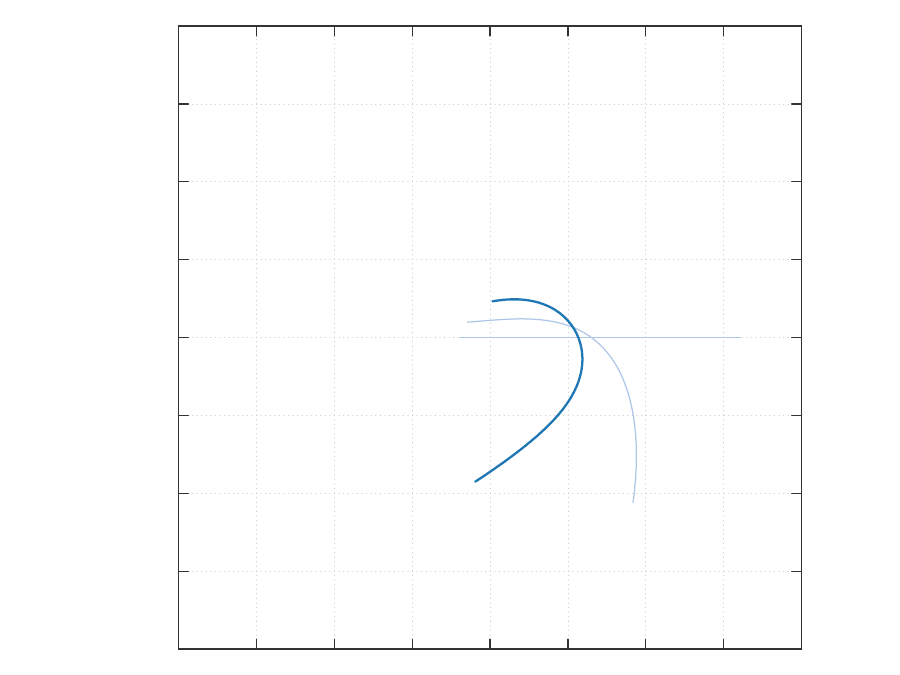}
  \caption{$t = t_1$}
\end{subfigure}
\hfill
\begin{subfigure}[b]{0.19\textwidth}
  \centering
  \includegraphics[width=\textwidth]{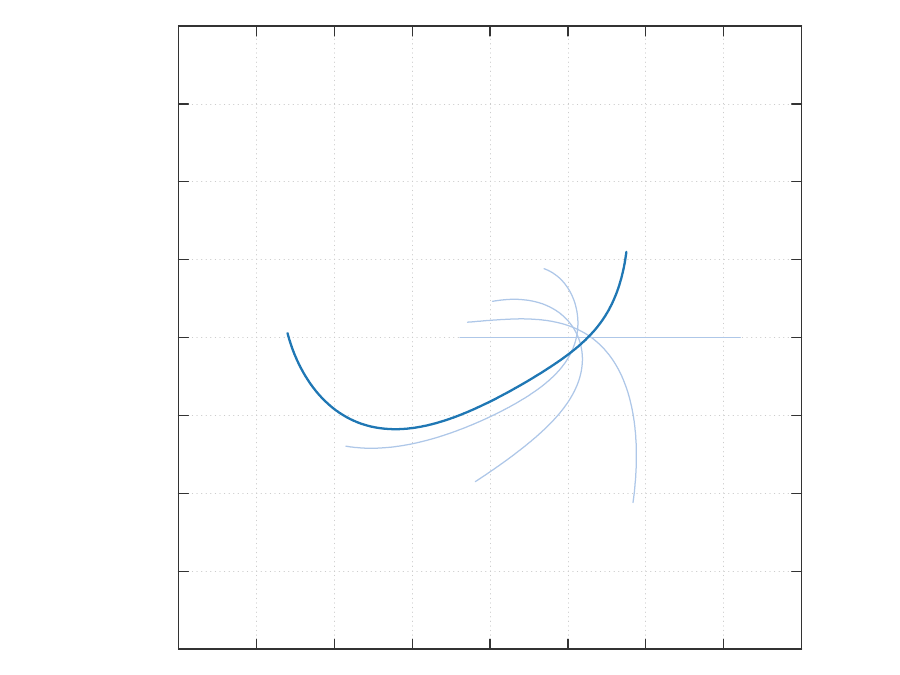}
  \caption{$t = t_2$}
\end{subfigure}
\hfill
\begin{subfigure}[b]{0.19\textwidth}
  \centering
  \includegraphics[width=\textwidth]{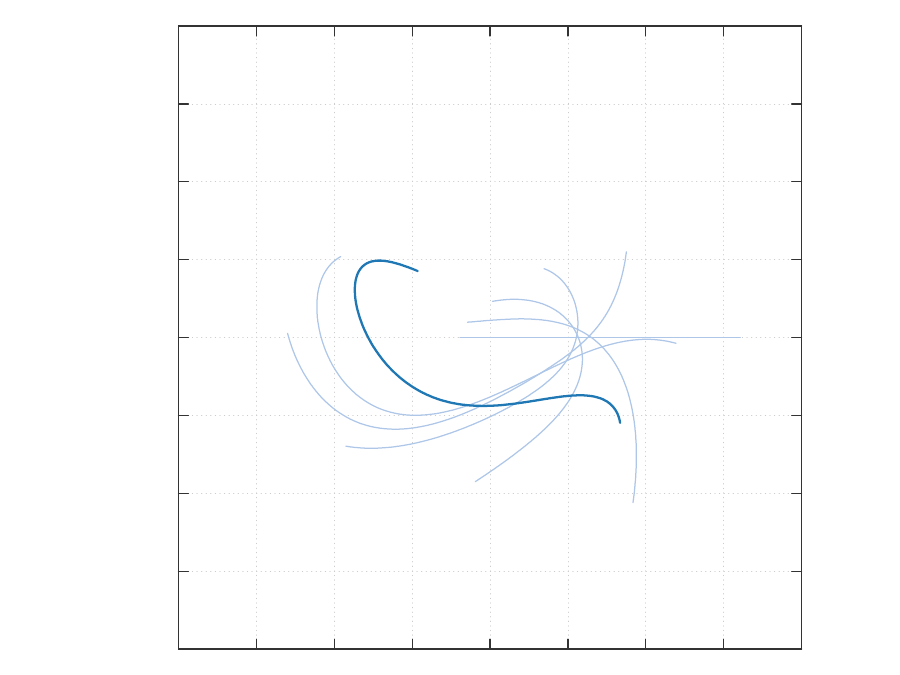}
  \caption{$t = t_3$}
\end{subfigure}
\hfill
\begin{subfigure}[b]{0.19\textwidth}
  \centering
  \includegraphics[width=\textwidth]{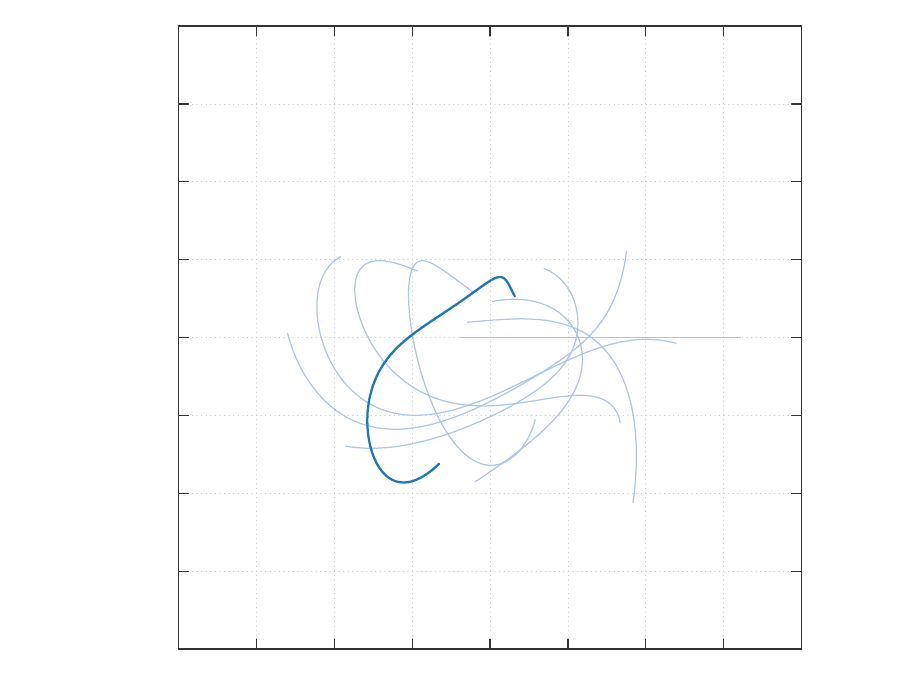}
  \caption{$t = t_4$}
\end{subfigure}

\caption{\textbf{Finite-dimensional dynamical trajectories across distinct energy regimes.} 
Phase-space projections in reduced dimensions (horizontal axis: field amplitude $\phi$, vertical axis: conjugate momentum $\dot{\phi}$) showing the reduced-order dynamics of the infinite-dimensional field system. 
\textbf{(a–e) Regime I} ($A = 0.2 \ll \phi_{v1}$): Trajectories confined within a localized low-dimensional manifold around a single vacuum. 
\textbf{(f–j) Regime II} ($A = 1.0 \approx \phi_{v1}$): Symmetric breather dynamics manifesting as structured low-dimensional orbits bridging the dual vacua (see Supplemental Movie 2 for spatio-temporal dynamics and Supplemental Movie 5 for phase-space motion). 
\textbf{(k–o) Regime III} ($A = 1.8 > \phi_{v1}$): Complex, high-dimensional phase mixing caused by barrier overflow, leading to spatial dispersion across modes (see Supplemental Movie 3 for spatio-temporal dynamics and Supplemental Movie 6 for phase-space motion). 
In each snapshot, dark blue curves indicate the present state, while light blue curves depict recent trajectories as a guide to the eye.
The horizontal axis denotes $ \phi \in [-2, 2]$ and the vertical axis represents conjugate moment $\pi = {\dot \phi} \in [-2, 2]$ for (a-e), $\phi \in [-2, 2]$ and $\pi = {\dot \phi}  \in [-0.3, 0,3]$ for (f-i), and $\phi \in [-4, 4]$ and $\pi = {\dot \phi}  \in [-8, 8]$ for (k-o). 
}
\label{fig:finite_dimensional_rep}
\end{figure*}

\section{\label{sec:results} Results and Discussion}

\subsection{Effective Curvature and Dynamical Mass Generation}
In standard quantum field theory, the physical mass of a scalar excitation is defined by the localized, static curvature around a vacuum minimum:
\begin{equation}
m_{\text{static}}^2 = \left. \frac{d^2 V(\phi)}{d\phi^2} \right|_{\phi = \phi_{v1}} = V''(\phi_{v1}) > 0,
\end{equation}
which inherently assumes small-amplitude oscillations confined near the ground state.

When considering non-perturbative configurations undergoing large-amplitude oscillations, the dynamic effective mass is governed by the time-averaged effective curvature sampled along the actual trajectory $\phi(t)$:
\begin{equation}
m_{\text{eff}}^2 \propto \langle V''(\phi(t)) \rangle_t = \frac{1}{T} \int_0^T V''(\phi(t)) \, dt.
\end{equation}
For the $Z_2$-symmetric potential $V(\phi) = \frac{1}{2}\mu^2\phi^2 + \frac{1}{4}\lambda\phi^4$ ($\mu^2 < 0$), the second derivative $V''(\phi) = \mu^2 + 3\lambda\phi^2$ exhibits distinct sign sectors partitioned by the inflection points $|\phi| = \phi_{v1}/\sqrt{3}$, as summarized in Table~\ref{tab:curvature_sectors}.

A field trajectory traversing both positive- and negative-curvature regions experiences dynamic cancellation between positive and negative contributions, providing an operator-theoretic mechanism for mass suppression without parameter fine-tuning (Fig.~\ref{fig:schematic_regimes}).
To validate this mechanism, Figs.~\ref{fig:field_profiles} and \ref{fig:finite_dimensional_rep} present the numerically computed spatio-temporal dynamics and their reduced-order finite-dimensional phase-space representations across Regimes I, II, and III, where $(\pm\sqrt{2}, 0)$ correspond to the uniform vacua.

\subsection{Dynamical Classification of Regimes}
The field behavior across different amplitude regimes is systematically governed by the interplay between the oscillation amplitude $A$ and the depth of the $Z_2$ potential landscape.

\subsubsection{Small Amplitude Limit ($A \ll \phi_{v1}$): Static $Z_2$ Symmetry Breaking}
When $A \ll \phi_{v1} = \sqrt{-\mu^2/\lambda}$, kinetic energy is insufficient to overcome the central barrier. 
The field undergoes small-amplitude localized oscillations confined within a single potential well [Figs.~\ref{fig:field_profiles}(a--e) and \ref{fig:finite_dimensional_rep}(a--e)].
In reduced phase space, the trajectory remains strictly confined to a low-dimensional manifold executing small orbits around $(+\sqrt{2}, 0)$. 
This recovers the conventional static picture of spontaneous symmetry breaking (SSB), generating a constant static mass $m_{\text{static}}^2 = V''(\phi_{v1})$.

\subsubsection{Intermediate Amplitude ($A \approx \phi_{v1}$): Dual Breathers}
As the amplitude approaches the barrier depth, the symmetric dual breather mode manifests as a spatio-temporal wave packet with synchronized, phase-inverted oscillations [Figs.~\ref{fig:field_profiles}(f--j) and \ref{fig:finite_dimensional_rep}(f--j)].
In reduced phase space, the system forms structured low-dimensional orbits bridging both vacuum sectors, where spatial domains oscillate around negative and positive vacua respectively.

\subsubsection{Large Amplitude ($A > \phi_{v1}$): The Heteroclinic (Majorana) Regime}
When $A$ exceeds the barrier height, the field overflows the central barrier at $\phi = 0$, exhibiting continuous large-scale spatial inversions [Figs.~\ref{fig:field_profiles}(k--o) and \ref{fig:finite_dimensional_rep}(k--o)].

Unlike localized single-well motion characterized by homoclinic-like bound trajectories, exceeding the threshold amplitude ($A > \phi_{v1}$) opens a global heteroclinic channel. This heteroclinic orbit continuously bridges the symmetric vacua ($\pm \phi_{v1}$), balancing residence times in positive-curvature wells ($V'' > 0$) and the negative-curvature peak ($V'' < 0$). This geometric balance naturally drives the effective mass suppression $\langle V''(\phi) \rangle_t \to 0$.
Furthermore, continuous flipping between parity sectors yields a vanishing time-averaged order parameter ($\langle \phi \rangle = 0$), realizing Majorana properties dynamically without introducing static axial backgrounds.

\subsection{Cancellation Mechanism with $\mu$, $\lambda$, and $A$ Dependence}
To quantitatively establish the predictability and scale independence of this framework, we formulate how the effective mass\footnote{Note: The evaluation of the numerical value of $m_{\mathrm{eff}}^2$ itself remains equivalent to that in the standard real/complex field framework through the background field expectation value $\langle \phi(t)^2 \rangle_t$. The primary role of the present geometric mechanism is not to introduce a new parameter scale, but to dynamically govern the chiral selection and the emergence of the self-conjugate Majorana nature on top of this effective mass.} depends on the potential parameters $\mu^2 (< 0)$, $\lambda$, and the initial amplitude $A$:
\begin{equation} \label{eq:m_eff_param}
m_{\text{eff}}^2 \equiv \langle V''(\phi(t)) \rangle_t = \mu^2 + 3\lambda \left\langle \phi(t)^2 \right\rangle_t.
\end{equation}
Scaling the field profile relative to the classical vacuum expectation value, $\phi_{v1} = \sqrt{-\mu^2/\lambda}$, recasts the normalized effective mass ratio into a dimensionless form:
\begin{equation} \label{eq:scaling_function}
\frac{m_{\text{eff}}^2}{|\mu|^2} = -1 + 3 \left\langle \left( \frac{\phi(t)}{\phi_{v1}} \right)^2 \right\rangle_t \equiv \mathcal{F}\left( \frac{A}{\phi_{v1}} \right),
\end{equation}
where $\mathcal{F}(y)$ represents a universal, dimensionless scaling function governed solely by the relative amplitude ratio $y \equiv A / \phi_{v1}$.

This formulation yields a clear separation between the bare dimensional energy scale $|\mu|$ and the geometric orbital factor $\mathcal{F}(y)$. 
Through the lens of this scaling function, conventional quantum field theory effectively fixes $\mathcal{F}(y) \equiv 2$, corresponding to the small-amplitude limit ($y \ll 1$), where trajectories are deeply trapped around a single vacuum well ($\langle (\phi / \phi_{v1})^2 \rangle_t \approx 1$) and recover the static mass $m_{\text{eff}}^2 \approx 2|\mu|^2$.

By contrast, promoting $\mathcal{F}(y)$ to a fully dynamical, amplitude-dependent orbital function opens the heteroclinic overflow regime ($y \gtrsim 1$), where the orbit traverses both positive- and negative-curvature sectors to drive $\mathcal{F}(y) \to 0^+$. 
Importantly, this universal function demonstrates that dynamic cancellation shifts the resolution of the hierarchy problem from static parameter fine-tuning of $\mu$ or $\lambda$ to the dynamic regime selection of the orbit $y = A / \phi_{v1}$. 
Near the heteroclinic threshold ($y \to 1^+$), the orbital period diverges logarithmically, naturally extending the time spent in the negative-curvature region ($V'' < 0$). 
This logarithmic scaling allows $\mathcal{F}(y)$ to achieve substantial order-of-magnitude suppressions down to sub-eV scales without requiring trans-Planckian or unnaturally small bare input masses. 
Consequently, the active neutrino mass scale emerges as a natural consequence of invariant manifold dynamics, where the extreme smallness reflects the geometric proximity to the heteroclinic boundary rather than a fine-tuned static mass parameter.

\subsection{Physical Implications for Neutrinos and Mass Eigenstates}
Applying this non-perturbative framework to neutrino physics, active neutrinos are identified with the large-amplitude heteroclinic regime ($A > \phi_{v1}$), providing geometric solutions to key phenomenological features:

\paragraph{Eigenstates and Mass-Squared Differences}
Each mass eigenstate $i \in \{1, 2, 3\}$ corresponds to a distinct heteroclinic trajectory with amplitude $A_i$ and frequency $\omega_i$, generating an independent effective mass $m_i^2 \propto \langle V''(\phi) \rangle_{t, i}$. Observed mass-squared differences $\Delta m_{ij}^2$ naturally emerge as nonlinear frequency shifts between these distinct topological orbits, while flavor mixing ($\nu_e \leftrightarrow \nu_\mu \leftrightarrow \nu_\tau$) represents dynamical phase-interference between coupled trajectories.

\paragraph{NG Counterparts and Residual Mass ($Z_2$ vs. $U(1)$)}
While continuous $U(1)$ symmetry breaking yields a massless Nambu-Goldstone mode along a flat manifold, the symmetric dual breather in discrete $Z_2$ systems serves as its dynamic counterpart, coupling opposite-phase excitations across the barrier to enforce masslessness ($\langle V'' \rangle_t \to 0$).
In a complex $U(1)$ system $\Phi(t) = \rho(t) e^{i\theta(t)}$, a conserved Noether charge $Q \propto \rho^2 \dot{\theta}$ generates a centrifugal barrier $\sim Q^2 / \rho^2$, preventing trajectories from dwelling at the origin ($\rho = 0$). 
This induces a subtle residence-time asymmetry, accounting for the tiny, non-zero mass scale of active neutrinos:
\begin{equation}
0 < m_{\text{eff}}^2 \propto \langle V''(\rho(t)) \rangle_t \ll m_{\text{static}}^2.
\end{equation}

\paragraph{Dynamic Mass Generation and Condensed Matter Duality}
Unlike QFT models relying on static Higgs VEV parameters, this paradigm aligns with QCD where hadronic mass predominantly originates from dynamic symmetry breaking. 
Furthermore, this heteroclinic barrier overflow shares a mathematical origin with 1D Ising-like spin systems~\cite{1966LiebMattis,1979Solyom,2003Giamarchi,1961Lieb,2001Kitaev}, where Jordan-Wigner transformations yield localized Majorana modes.
More broadly, such spatio-temporal localization and dynamic phase coexistence mirror universal pattern formation mechanisms observed in non-equilibrium and nonlinear physical systems~\cite{93Cross,15Ball,20Nabika}, demonstrating that such topological mechanisms extend naturally from continuous classical fields to macroscopic domain structures.

\subsection{Dynamical Emergence of Left-Handed Chirality and Weyl Limit}
Beyond effective mass suppression and Majorana self-conjugacy, the phase-space geometry of the heteroclinic channel offers a fundamental explanation for the observed left-handed chirality of active neutrinos. While our underlying order parameter is governed by a real scalar field $\phi(t,x)$, its non-perturbative spatial inversion dynamics in spacetime can be mapped onto a real Majorana spinor representation via generalized Clifford-algebraic projections. 
We propose the following effective chiral description to model the phase-space rectification (cf. finite-dimensional representation)  and mode decoupling.
By transforming this real representation into the chiral basis, we consider a micro-scale system with two coupled chiral components $(\chi_L, \chi_R)$ interacting through the dynamic effective mass field $m_{\mathrm{eff}}(t)$ generated along the heteroclinic orbit:
$$i \bar{\sigma}^\mu \partial_\mu \chi_L = m_{\mathrm{eff}}(t) \chi_R,$$
$$i \sigma^\mu \partial_\mu \chi_R = m_{\mathrm{eff}}(t) \chi_L.$$
Here, under the self-adjoint constraint of the underlying real scalar field, these components naturally satisfy the Majorana self-conjugacy relation ($\chi_R \sim \chi_L^c$).

Near the heteroclinic threshold ($A \ge \phi_{v1}$), the continuous barrier-overflow dynamics introduces a highly nonlinear phase factor $\theta(t,x) = kx - \omega t + \phi_{\mathrm{nl}}(t)$, where $\phi_{\mathrm{nl}}(t)$ directly reflects the periodic spatial inversions across the central barrier. 
When this nonlinear phase interferes with the spatial wavenumber $k$, the mode coupling exhibits a topological rectification effect. 
Due to the inherent directional orientation of the heteroclinic flow along the spatial wavenumber vector $k$, this nonlinear interference generates a chirality-selective dynamical damping.
Specifically, we model this rectification phenomenologically by assuming that the effective dissipation acts primarily on the positive-helicity subspace. 
The phase-space coupling between the nonlinear inversion phase $\phi_{\mathrm{nl}}(t)$ and the spatial wave vector motivates an effective dissipation operator aligned with the positive-helicity projection, $\frac{1}{2}(I + \vec{\sigma} \cdot \hat{k}) = P_R = \frac{1+\gamma_5}{2}$, in the ultra-relativistic limit.
Formally, using the chiral projection operators $P_L = \frac{1-\gamma_5}{2} = \begin{pmatrix} 1 & 0 \\ 0 & 0 \end{pmatrix}$ and $P_R = \frac{1+\gamma_5}{2} = \begin{pmatrix} 0 & 0 \\ 0 & 1 \end{pmatrix}$, the effective dissipation acts asymmetrically on the chiral doublet $\Psi = \begin{pmatrix} \chi_L \\ \chi_R \end{pmatrix}$:
$$\frac{d}{dt}\begin{pmatrix} \chi_L \\ \chi_R \end{pmatrix} + \Gamma(k,A) P_R \begin{pmatrix} \chi_L \\ \chi_R \end{pmatrix} = \begin{pmatrix} \Theta_L(m_{\mathrm{eff}}\chi_R) \\ \Theta_R(m_{\mathrm{eff}}\chi_L) \end{pmatrix},$$
which explicitly expands to:
$$\frac{d\chi_R}{dt} + \Gamma(k,A)\chi_R = \Theta_R(m_{\mathrm{eff}}\chi_L),$$
$$\frac{d\chi_L}{dt} = \Theta_L(m_{\mathrm{eff}}\chi_R),$$
where $\Gamma(k,A) > 0$ denotes the dynamical dissipation rate acting exclusively on the right-handed sector along the unstable manifold. 
Within this effective description, the absence of a damping coefficient in the left-handed sector follows immediately from the projected dynamics\footnote{Note: Unlike heteroclinic channels connecting distinct vacuum expectation values, homoclinic orbits ($Q=0$) possess symmetric phase-space trajectories that preserve parity, thereby precluding such chirality-selective dynamical rectification.}.
Consequently, the right-handed sector $\chi_R$ rapidly decouples or dissipates from the low-energy asymptotic trajectory ($\chi_R \to 0$), leaving $\chi_L$ as the sole surviving, norm-preserving component. In this asymptotic limit, the surviving left-handed component $\chi_L$ satisfies the decoupled, scale-invariant Weyl equation:
$$i \bar{\sigma}^\mu \partial_\mu \chi_L = 0,$$
where $\bar{\sigma}^\mu = (I, -\vec{\sigma})$ represents the Pauli matrix vector. This derivation demonstrates that left-handed chirality is not merely a hand-inserted gauge-theoretic postulate, but an emergent, dynamic property resulting from phase-space rectification and right-handed mode decoupling along the heteroclinic invariant manifold.

\section{\label{sec:conclusion} Conclusion}
In this paper, we have established a dynamic, nonperturbative framework for $Z_2$ spontaneous symmetry breaking in scalar field theory, redefining the conventional static view as a low-energy, small-amplitude limit ($A \ll \phi_{v1}$). 
Indeed, at sufficiently high energies, the dynamics floods the potential landscape, making the SSB-induced double-well structure irrelevant.
By formulating symmetry breaking through operator-theoretic spectral branching, we demonstrated that the real-valued nonlinear Klein-Gordon equation naturally accommodates localized, phase-inverted bound states—termed the ``symmetric dual breather mode.'' This state constitutes a robust, persistently stable nonperturbative excitation within the real scalar field framework, filling a critical blind spot in nonperturbative field dynamics.

Utilizing a $t$-dependent finite-dimensional representation, we classified the system's dynamical spectrum across three distinct energy regimes. As the field amplitude increases into the heteroclinic regime ($A > \phi_{v1}$), the field dynamically overflows the central potential barrier, giving rise to a vanishing time-averaged order parameter ($\langle \phi \rangle = 0$) and a cancellation mechanism for effective mass generation ($m^2_{\text{eff}} \propto \langle V''(\phi) \rangle_t \to 0$). Applied to neutrino physics, this framework naturally reproduces remarkably small neutrino masses, mass-squared differences, flavor mixing, and emergent Majorana properties as topological limit cycles in reduced phase space, while guaranteeing observable CPT invariance through orbital time-averaging. Furthermore, generalizing from the real $Z_2$ limit to the complex $U(1)$ plane reveals that dynamic centrifugal forces naturally account for the tiny residual mass scale of active neutrinos.

Crucially, the symmetric dual breather mode does not represent a transient numerical artifact, but reflects the existence of a robust, parameter-dependent invariant manifold in phase space.
Along this manifold, the field undergoes a heteroclinic motion bridging the two symmetric vacuum sectors without dissipative relaxation.
Near the heteroclinic threshold, the logarithmic divergence of the orbital period naturally extends residence times in the negative-curvature region, allowing the universal scaling function $\mathcal{F}(y) \equiv m^2_{\text{eff}}/|\mu|^2$ to achieve extreme mass suppressions down to sub-eV scales.

This geometric constraint shifts the fundamental paradigm of spontaneous symmetry breaking from static parameter fine-tuning to invariant manifold dynamics.
Here, the purified real-valued dynamics provides a remarkable structural link: by traversing discrete parity sectors continuously rather than through a discontinuous first-order phase transition, this topological duality to condensed matter systems naturally drives the dynamical realization of Majorana properties.

Beyond particle masses, this conceptual transition from static parameters to dynamic orbital-averaging holds broad potential for addressing extreme scale hierarchies across fundamental physics, including the cosmological constant problem and gravitational scale disparities. 
In summary, this non-perturbative operator-theoretic framework establishes that the extreme suppression of active neutrino masses, their Majorana self-conjugacy, and the dynamical selection of left-handed chirality are described within an effective phenomenological framework based on the invariant manifold geometry of heteroclinic dynamics

\begin{acknowledgments}
This research was partially supported by Osaka University of Economics and Law.
\end{acknowledgments}



\nocite{*}

\end{document}